\documentclass[a4paper,fleqn,usenatbib]{mnras}

\usepackage{ae,aecompl}

\usepackage{graphicx}	
\usepackage{amsmath}	
\usepackage{amssymb}	

\newcommand\ams{The Annals of Mathematical Statistics}

\title[New SMC cluster candidates]{The VMC Survey - XXI.  New star cluster candidates discovered from infrared
photometry in the Small Magellanic Cloud{\thanks{Based on observations collected at the European Organisation for Astronomical Research in the Southern Hemisphere under ESO programme(s) 179.B-2003.}}}

\author[A.E. Piatti et al.]{
Andr\'es E. Piatti$^{1,2}$\thanks{E-mail: andres@oac.unc.edu.ar (AEP)}, Valentin D. 
Ivanov$^{3}$, Stefano Rubele$^{4,11}$, Marcella Marconi$^{5}$,  
\newauthor Vincenzo Ripepi$^{5}$, Maria-Rosa L. Cioni$^{6,7,8}$, Joana M. Oliveira$^{9}$ 
\newauthor and Kenji Bekki$^{10}$\\
$^{1}$Observatorio Astron\'omico, Universidad Nacional de C\'ordoba, Laprida 854, 5000, 
C\'ordoba, Argentina\\
$^{2}$Consejo Nacional de Investigaciones Cient\'{\i}ficas y T\'ecnicas, Av. Rivadavia 1917, 
C1033AAJ, Buenos Aires, Argentina\\
$^{3}$European Southern Observatory, Karl-Schwarzschild-Str. 2, 85748 Garching bei M\"nchen, 
Germany\\
$^{4}$ INAF, Osservatorio Astronomico di Padova, vicolo dell’Osservatorio 5, I-35122 Padova, Italy\\
$^{5}$ INAF-Osservatorio Astronomico di Capodimonte, via Moiariello 16, 80131, Naples, 
Italy\\
$^{6}$ Universit\"at Potsdam, Institut f\"ur Physik und Astronomie, Karl-Liebknecht-Str. 24/25, 14476 Potsdam, Germany\\
$^7$ Leibnitz-Institut f\"ur Astrophysik Potsdam, An der Sternwarte 16, 14482 Potsdam, Germany\\
$^8$ University of Hertfordshire, Physics Astronomy and Mathematics, College Lane, Hatfield AL10 9AB, United Kingdom\\
$^{9}$ Lennard-Jones Laboratories, School of Physical and Geographical Sciences, Keele University, ST5 5BG, UK\\
$^{10}$ ICRAR, University of Western Australia, 35 Stirling Hwy, Crawley WA 6009, Australia\\
$^{11}$ Dipartimento di Fisica e Astronomia, Universita' di Padova, Vicolo
> dell'Osservatorio 2, I-35122 Padova, Italy\\
}

\date{Accepted XXX. Received YYY; in original form ZZZ}

\pubyear{2015}

\begin{document}
\label{firstpage}
\pagerange{\pageref{firstpage}--\pageref{lastpage}}
\maketitle

\begin{abstract}
We report the first search for new star clusters performed using the VISTA 
near-infrared $YJK_s$ Magellanic Clouds survey (VMC) data sets. We chose a pilot field 
of $\sim$ 0.4 deg$^2$ located in the South-West of the Small Magelllanic Cloud
(SMC) bar, where the star field is among the densest and highest reddened region 
in the galaxy. In order to devise an appropriate automatic procedure we made use of
dimensions and stellar densities observed in the VMC data sets of the 
known clusters in this area. We executed different kernel density
estimations over a sample of more than 358000 stars with magnitudes measured in the
three $YJK_s$ filters. We analysed the new cluster candidates
whose colour-magnitude diagrams (CMDs), cleaned from field star contamination, 
were used to assess the clusters' reality and estimate reddenings and ages of the genuine systems. 
As a result 38 objects ($\approx$ a 55$\%$ increase in the known star clusters
located in the surveyed field) of 0.15 - 0.40 arcmin (2.6 - 7.0 pc) in 
radius resulted to have near-infrared CMD features which resemble those of star clusters 
of young to moderate intermediate age (log($t$ yr$^{-1}$) $\sim$ 7.5-9.0). Most of the new star cluster candidates are hardly recognizable
in optical images without the help of a sound star field decontaminated CMD analysis.
For highly reddened star cluster candidates ($E(B-V)$ $\ge$ 0.6 mag) the VMC data sets were necessary
in order to recognize them.

\end{abstract}




\section{Introduction}

The Magellanic Clouds (MCs) offer us a unique opportunity to study
star clusters that have been formed in distinct
environments -- including age, metallicity, overall gravitational
potential, and formation history -- in comparison with the star cluster system of the
Milky Way, to which we are most familiar with 
\citep[see][to trace the developments in the field]{vdb91,dga06,baetal13}.

Star clusters in the Large and Small Magellanic Clouds (L/SMC) are spread over many 
hundreds of square degrees. In comparison, this area is much smaller than the entire sky
that must be surveyed for a complete census of the Milky Way's
star clusters, but the distance to the MCs and the crowding
from the dense MC field population, make the MC star cluster census
a challenging task, subjected to conflicting constraints --
it requires deep high angular resolution wide-field imaging.

Today hundreds of SMC star clusters are known \citep{bd00,betal08}. 
The first star cluster searches in the
galaxy were carried out nearly a century ago.
\citet{sw25} compiled a catalog of non-stellar objects,
and later \citet{k56} realized
that many of them were star clusters.\citet{m35} 
reported seven star clusters in the SMC outskirts. Most of the
objects in these early lists are prominent enough to have NGC
entries.

\citet{k56} and \citet{l58} performed special searches
aimed at identifying star clusters in the SMC, and indeed they recognized 69
and 116 objects, respectively. Further lists of star clusters were
published by \citet[][18 objects]{wg71}, \citet[][86 objects]{hw74}, and 
\citet[][168 objects]{b76}. The
compilation of \citet{hw77} contained 220 star clusters.
\citet{h86} reported 233 new star clusters, and the major effort by
\citet{bs95} brought the total number of known SMC
clusters to 554 -- the apex of the star cluster searches based on
photographic plates.

The next major step was carried out by \break
\citet{pieetal98},
who utilized the wide field digital imaging in the optical
carried out by the Optical Gravitational Lensing Experiment 2 \citep[OGLE2][]{p86,udalskietal97} 
to search for star clusters within $\sim$2.4\,deg$^2$ near to the SMC centre. 
They reported 238 star clusters, 72 of which were new identifications. Most
notably, they applied for the first time an automated
technique based on surface density maps, following \citet{zaritskyetal97}
 who used that procedure to search for star clusters in the LMC. Unlike previous 
efforts that were in 
effect a simple visual inspection of photographic plates or prints, this method
is objective, and is able to assign probabilities to candidates,
based on how much they exceed the statistical fluctuations over the
background level. Nevertheless, visual inspection of the candidates and their
colour-magnitude diagrams (CMDs) are still desirable, because of star clusters
embedded in gas clouds might exist.
Interestingly, automated methods are much more commonly applied
to identify Milky Way's star clusters: 
\citet{ivanovetal02,borissovaetal03,merceretal05,froebrichetal07,koposovetal08}. 

\citet{bd00} sumarized the state of the SMC star cluster
census at that time. They listed 719 star clusters in total. This
number included young star clusters with emission, and loose
systems. Later, \citet{betal08} updates the SMC star cluster
catalogue. A finer re-classification of some objects into
star clusters, associations, and shells reduced the number of
SMC star clusters to 498. Their properties (e.g., size, age) vary within wide
limits, but this discussion is beyond the scope of this
paper.

This paper is organised as follows. Section 2 provides an overview of
the data sets used in this study and describes the strategy to identify new star 
clusters. In Section 3
we present the new identified star cluster candidates, the procedure to clean their CMDs from 
the contamination of field stars and the use of isochrone to determine the 
clusters' best-fitting physical parameters. In the same Section we explore the
star cluster observed features in both the optical
and the near-infrared wavelength ranges. We present our main conclusions in
Section 4.

\section{strategy for searching new star clusters}

In this paper we devise a procedure to search for new star clusters in the SMC with the
aim of applying it to the whole VISTA\footnote{Visible and Infrared Survey Telescope 
for Astronomy} near-infrared $YJK_s$ survey of the Magellanic Clouds system 
\cite[VMC][]{cetal11}. VMC is an ESO (European Southern
Observatory) public survey that is carried out with the VIRCAM (VISTA InfraRed
Camera) instrument \citep{daltonetal2006} on the ESO VISTA telescope 
\citep{eetal06}. The main goals of the survey are to reconstruct the star-formation 
history (SFH) and its spatial variation, as well as to infer an accurate 3D map of the 
entire Magellanic system.  The VMC survey strategy involves repeated observations of tiles across the Magellanic system, where one tile covers uniformly
an area of $\sim$ 1.5 deg$^2$, as a result of the mosaic of six paw-print
images, in a given waveband with 3 epochs at $Y$ and $J$, and
12 epochs at $K_s$. Individual epochs have exposure times of 800 s
($Y$ and $J$) and 750 s ($K_s$). The average quality of the VMC data
analysed here corresponds to 0.34$\arcsec$ pixel size and 0.90$\arcsec$ FWHM.

We chose a 36$\times$36 arcmin$^2$ pilot field located in the South-West part of the 
SMC bar (see inset chart in Fig. 1) for which we have available images and
photometry in the Washington $CT1$ photometric system \citep{p12a}.
These optical images and photometry were used for comparison purposes, i.e., in order to know
whether possible new star clusters from the VMC survey were not discovered until now 
because of the high reddening, the limiting magnitude, the clusters' sizes and/or the cataloguing
technique. The pilot field is likewise one of the densest fields in the galaxy 
and presents noticeable variations in the stellar density and in the interstellar reddening.

\begin{figure}
	\includegraphics[width=\columnwidth]{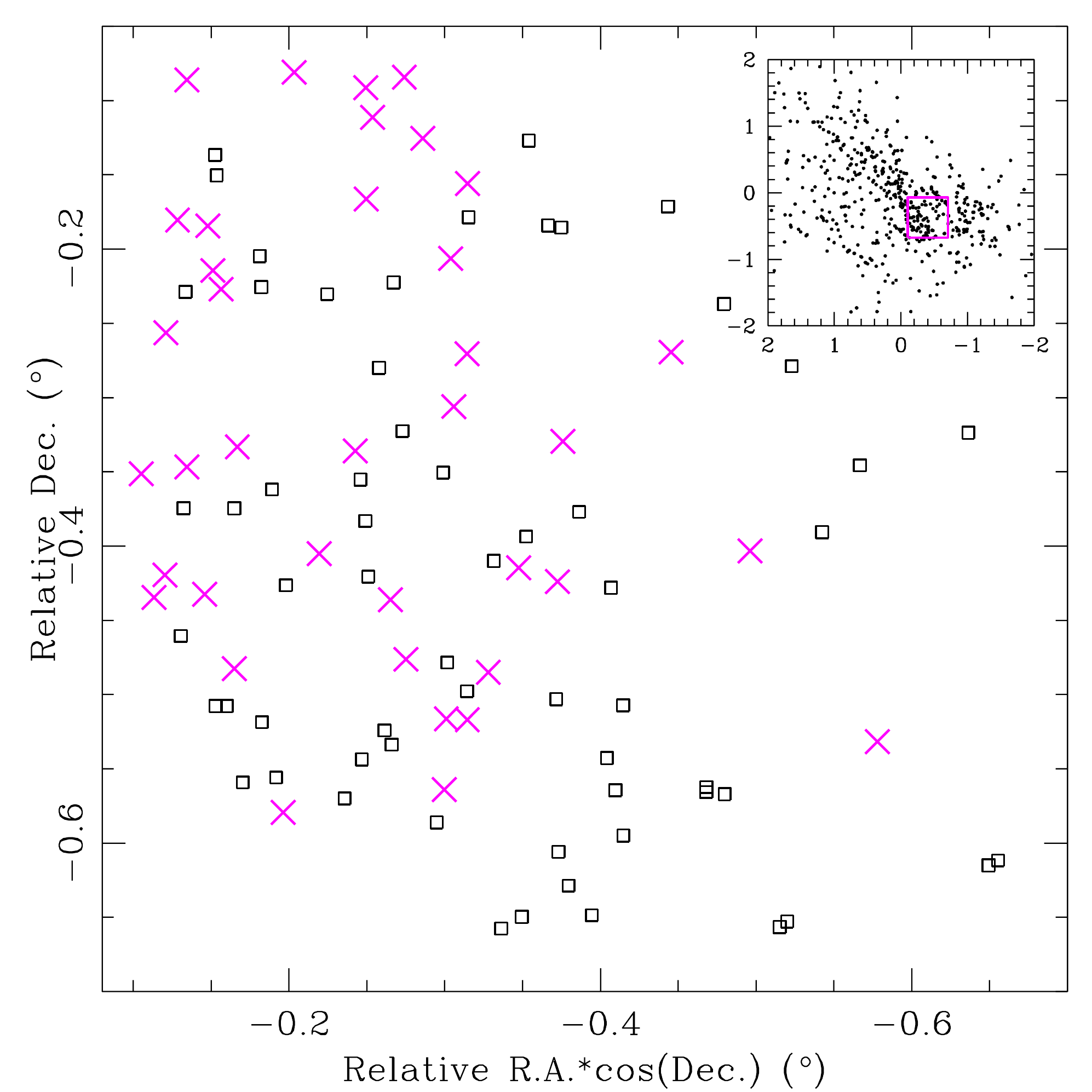}
    \caption{Pilot region chosen to search for new star clusters. Open black squares and
magenta crosses represent previoulsy catalogued and new star clusters, respectively. The 
inset panel depicts the spatial distribution of the catalogued SMC star clusters. A 
thick magenta-coloured open box delimits the studied SMC field. We used as the SMC centre: 
RA = 00h 52m 45s, Dec = $-$72$\degr$ 49$\arcmin$ 43$\arcsec$ (J2000) \citep{cetal01}.}
    \label{fig:fig1}
\end{figure}

The Washington $CT_1$ images of the seleted field were obtained at the Cerro Tololo 
Inter-American Observatory 4-m Blanco telescope with the Mosaic II camera attached 
(a 8K$\times$8K CCD detector array) and are available at the National Optical Astronomy 
Observatory (NOAO) Science Data Management 
Archives\footnote{http://www.noao.edu/sdm/archives.php.}.
The 50 per cent completeness level of the resulting photometry is located at a $T_1$ 
magnitude and a $C-T_1$ colour corresponding to the Main Sequence (MS) turnoff of a 
stellar population with an age $\ga$ 10 Gyr. This photometric data set was used
to estimate fundamental parameters of star clusters \citep{p11b,metal14}, to trace the 
field age-metallicity relationship  \citep{p12a}, to develop a procedure to clean the 
clusters' CMDs from the unavoidable star field contamination 
\citep{pb12}, among others.

From the VMC survey, we used the region of the tile SMC 4$\_$3 which fully contains the selected field.
All data used in this work were produced as in the SFH study performed by \citet{retal15}, where
data acquisition and reduction, point-spread-function photometry and artificial star 
tests for completeness assessment are extensively described. Those VMC tile SMC 4$\_$3 data 
(88\% of completion) were used to trace a reddening map throughout the tile, to investigate the SMC depth along the 
line-of-sight, to build a look back time star formation rate and a mass tomography
\citep{retal15}. 
The reduced data
for this tile used here (100\% of completion) will be released to the community together with the data 
for other tiles as part of data release \#5 which is planned for 2016. 
All sources detected are not saturated, so that there are no stars missed at the bright part of the CMDs that may influence the subsequent star cluster analysis, contrarily to what was suggested by \citet{bicaetal2015}.
The 50 per cent completeness level reached in the $K_s$ mag and the $Y-K_s$ colour corresponds 
to the MS 
turnoff of a stellar population with an age of $\sim$ 6 Gyr old. 

We built a list of 68 star clusters located in the region from the the catalogue of 
\citet{betal08}. 
These catalogued star clusters are not always straightforward to identify in
 deep VMC tile images. This is because they were originally identified 
from optical images (e.g. from the Digitized Sky Survey, DSS, 
images) which can look rather different compared with their appearance at 
near-infrared wavelengths. Furthermore, they were identified using images with
different spatial resolutions and depth than the VMC ones.
Consequently, some misidentification might occur. For instance, single relatively bright 
stars might look like  an unresolved compact star cluster in images of lower spatial resolution, 
or unresolved background galaxies could be mistaken for small star clusters in shallower 
images. Fig. 1 depicts with
open squares the spatial distribution of star clusters catalogued by \citet{betal08} in 
the selected 
region. Because of the higher surface brightness of the background and the
more populous nature of this SMC bar region, star clusters stand out less, and incompleteness
effects are expected to be more important than in the outer regions of the galaxy.

We first overplotted the positions of the catalogued star clusters on the deepest Washington
$T_1$ and VMC $K_s$ images, and visually recognized them in both the optical and 
near-infrared regimes. Note that the main aim of this task is to confirm the given star cluster 
coordinates in order to properly count the number of stars inside the clusters' radii 
from the VMC PSF photometry. Fig. 2 shows the resulting mean stellar densities as 
a function of the star clusters' radii -taken from \citet{betal08}- for the catalogued star 
clusters in the surveyed region (open squares). The counts were carried out by considering only 
stars with magnitude measures in the three $YJK_s$ filters.
The range of values of both quantities -radii and stellar densities- were used as input 
information in the procedure of search for star clusters using the VMC data set. 

\begin{figure}
	\includegraphics[width=\columnwidth]{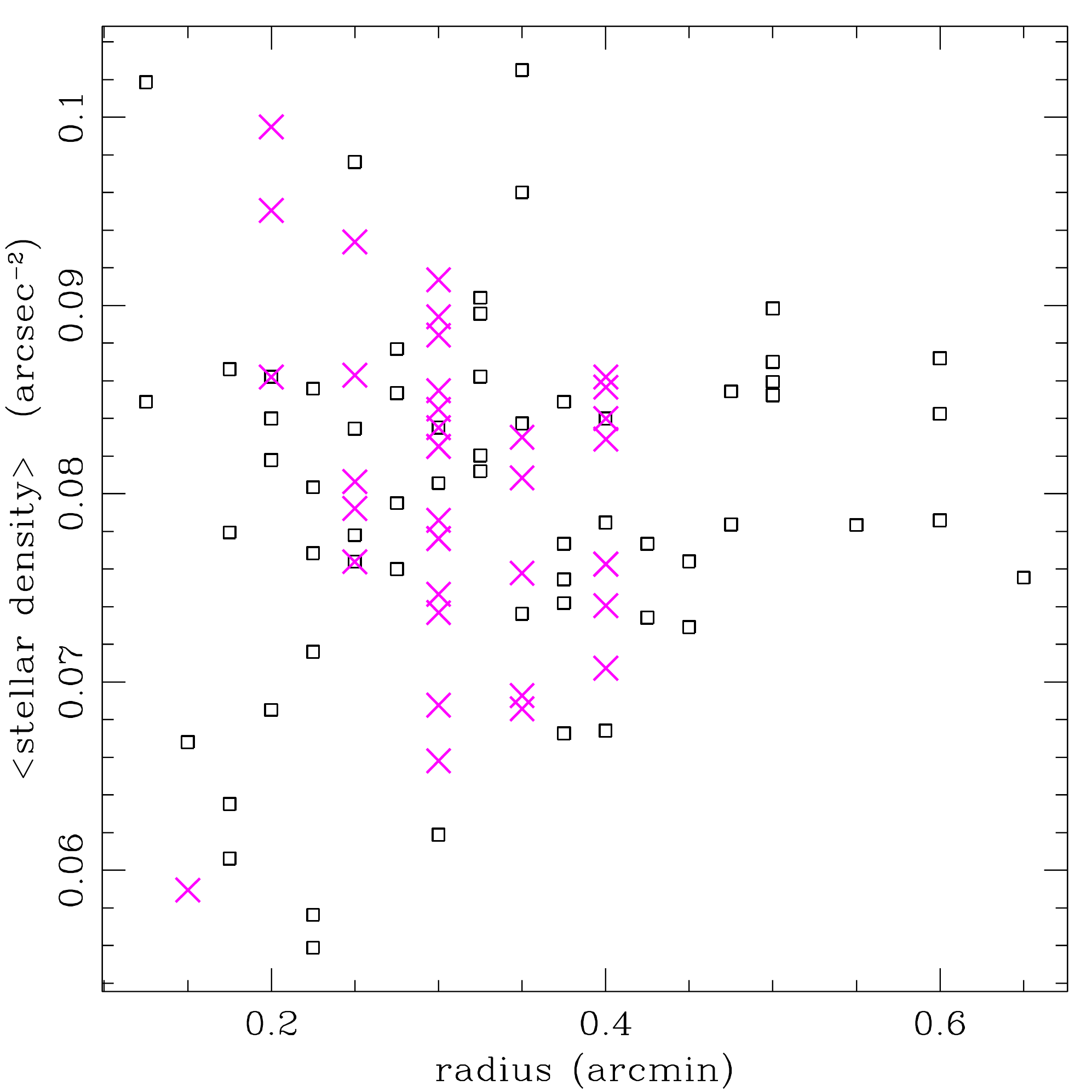}
    \caption{Relationship between the stellar density and the cluster radius in the studied 
SMC field. The typical background stellar density ranges between 0.008 and 0.012 arcsec$^{-2}$. Open black squares and magenta crosses represent previoulsy catalogued 
and new star cluster candidates, respectively.}
    \label{fig:fig2}
\end{figure}

The search was performed by employing AstroML\footnote{http://www.astroml.org/index.html} routines \citep[][and reference therein for a 
detail description of the complete AstroML package and user's Manual]{astroml}, a machine learning
and data mining for Astronomy package. AstroML is a Python module for machine learning and data mining built on {\it numpy, scipy, scikit-learn, matplotlib,} and {\it astropy}, and distributed under the 3-clause BSD license. It contains a growing library of statistical and machine learning routines for analysing astronomical data in Python, loaders for several open astronomical datasets, and a large suite of examples of analysing and visualizing astronomical datasets.
The goal of AstroML is to provide a community repository for fast Python implementations of common tools and routines used for statistical data analysis in astronomy and astrophysics, to provide a uniform and easy-to-use interface to freely available astronomical data sets.

We used two different  Kernel Density Estimators 
(KDEs) from six available in AstroML, namely, {\it Gaussian} and {\it tophat}, and three different bandwidths (the FWHM/2 of the KDE) for each KDE 
of  0.23, 0.45 and 0.68 arcmin, respectively, thus uniformly sampling the range of known star clusters' radii (see Fig. 2). A KDE is a non-parametric density estimation technique which alleviates the problem of the histogram dependence on 
the bin size and the end points of bins by centering a function (kernel) to each data point and then add them up \citep{r56}. The result is a continuous density
distribution - instead of a not smooth blocky histogram - that allows to extrat the finest structures of it. In our case, 
we generated a stellar density surface over the studied region. 
In total, we run six different kernel overdensity searches
on a sample of 358578 stars with positions and magnitudes measured in the three $YJK_s$ 
filters. From the total number of overdensities detected  respect to
their local backgrounds per search, we imposed a cut off density of 
0.05 arcsec$^{-2}$ in order to keep only star cluster candidades within the density range 
seen in Fig. 2 for the known star clusters.  The typical background
stellar density ranges between 0.008 and 0.012 arcsec$^{-2}$, so that we only keep
overdensities at least $\sim$ 5 times higher. Indeed, we identified all the 68 known
clusters from the VMC data set, which resulted to be overdensities higher than 0.05 arcsec$^{-2}$
(see Fig. 2). Note that the performance of the technique in identifying star clusters depends on
the chosen density cut off, bandwidths and KDE functions. Based on the selection criteria mentioned above, we used the most suitable ones in order to identify the 68 known star clusters and
new cluster candidates as well. We merged the resulting lists, avoiding repeated overdensities from different 
runs. We also discarded the recovered 68 known star clusters and visually inspected the star cluster candidates on the deepest $K_s$ image. This is a conservative exploration,  since star clusters 
embedded in clouds could exist and no overdensity would come out from this search \citep{romitaetal2016}.  
We finally got 143 star cluster candidates.

\section{results and discussion}

The new 143 apparent stellar concentrations found from the VMC data set do not necessarily assure us that 
they constitute physical systems. 
\citet{pb12} showed evidence that some SMC candidate star clusters are in fact not 
genuine physical systems. According to the authors, nearly 10 per cent of their studied 
star cluster sample resulted in possible non-clusters. This does not seem to be a significant 
percentage of the catalogued star clusters. Indeed, they used the spatial distribution of possible non-clusters in the SMC to statistically decontaminate that of the SMC star cluster system. 
By assuming that the area covered by them represents an unbiased subsample of the SMC as a 
whole and by using an elliptical framework centred on the SMC centre, they found that there 
is no significant difference between the expected and observed star cluster spatial distributions. 
However, a difference at a 2$\sigma$ level would become visible between $a$ $\approx$
0$\degr$.3 and 1$\degr$.32  (the semi-major axis of an ellipse parallel to the SMC bar and 
with $b/a$ = 1/2), if the possible non-clusters were increased up to 20 per cent. 
For the 143 new star cluster candidates, we used their CMDs to infer the existence of genuine 
star clusters.

When dealing with small objects or sparse star clusters projected or immersed
in crowded star fields, as those star clusters located in the inner regions of the SMC, 
a simple circular CMD extraction around the star cluster centre could lead to
a wrong conclusion, since the CMDs are obviously composed of stars of different stellar 
populations \citep{p12a}. Consequently, it is hardly possible to assess whether bright 
and young MSs or subgiant and red giant branches trace the fiducial star cluster features. 
In order to statistically  clean the star cluster CMDs from the unavoidable field contamination we
applied a procedure developed by \citet{pb12}, which has proved to be useful in previous
papers, among them, from the VMC data sets \citep[see, e.g.][]{petal14b,petal15a,petal15b}.
In short, the star field cleaning relies on the comparison of each
of four previously defined field CMDs to the cluster CMD and subtracted from the latter
a representative field CMD in terms of stellar density, luminosity function, and colour
distribution. This was done by comparing the numbers of stars counted in boxes distributed 
in a similar manner throughout all CMDs. The boxes were allowed to vary in size
and position throughout the CMDs in order to meaningfully represent the actual distribution 
of field stars.

Since we repeated this task for each of the four field CMDs, we could assign a membership 
probability to each star in the cluster CMD. This was done by counting the number of times
a star remained unsubtracted in the four cleaned cluster CMDs and by subsequently dividing 
this number by 4. Thus, we distinguished field populations projected on to the star cluster area, 
i.e. those stars with a probability $P \le$ 25 per cent, stars that could equally
likely be associated with either the field or the object of interest ($P =$ 50 per cent), 
and stars that are predominantly found in the cleaned star cluster CMDs ($P \ge$ 75 per cent) rather 
than in the field star CMDs. Statistically speaking, a certain amount of cleaning
residuals is expected, which depends on the degree of variability of the stellar density,
luminosity function and colour distribution of the field stars.
We employed this field star decontamination procedure to 
clean circular areas of 1.2 arcmin in radius around the central coordinates of the 143 new star 
cluster candidates using both Washington and VMC data sets. Here we took advantage
of the Washington CMDs for comparison purposes, thus allowing us to know from the optical regime
how cluster candidates identified from the near-infrared behave. Note that we do not use
the Washington photometry to infer the object reality.
As a result 38 objects
turned out to have cleaned near-infrared CMDs with features (stars with $P \ge$ 50 per cent) which resemble those of star clusters of young to 
moderate intermediate age (log($t$ yr$^{-1}$) $\sim$ 7.5-9.0). Cleaned CMDs without any detectable trace of star cluster sequences were discarded.

The star cluster radii used to extract their 
CMDs were taken from a visual inspection of the deepest $K_s$ image, and were meant to 
 allow us to meaningfully define
the clusters' fiducial sequences in their CMDs. The number of new star cluster
candidates represents
about a 55$\%$ increase in the number of known star clusters located in the surveyed region, which is a
significant percentage in terms of the currently debateable issues about the star cluster formation 
rate, the effectiveness of star cluster dissolution processes, etc 
\citep[][and references therein]{getal11,p14b,chandaretal2015}. 
Such a number could
be still higher if we carried out a search not as conservative as that developed in this
paper, i.e., not constrained to a range of stellar densities \citep[e.g. embedded star clusters][]{romitaetal2016}.
Note that we do not aim at
building a complete list of star clusters, but rather to devise a method useful to identify
star clusters making use of KDEs and, in turn, to introduce the first star cluster
candidates discovered
in the SMC based on such a procedure. It would be worth to apply the present analysis tools 
to search for faint poorly populated star clusters in the Magellanic
Clouds using deep images obtained by, for example, 8-m class telescopes.

Figures 1 and 2 show the positions in the galaxy of these new star cluster candidates and the 
relationship between their stellar densities and dimensions, respectively. We have highlighted 
them with magenta-coloured crosses. Note that they do not reveal any particular spatial
distribution pattern, but spread over the surveyed field. Likewise, their observed stellar
densities are within the range of those of known star clusters in the region.
This is a expected result, because of the imposed  cut off density limit and the
KDE's bandwidths used. The identification of new star cluster candidates shows that the automatic search 
performed here has some advantages over those previous carried out by visually inspecting 
photographic plates \citep[e.g.][]{bs95} or by other automatic algorithmic searches 
\citep[e.g.][]{pieetal98}. In Figure A1 of the Appendix we present 2$\times$2 
arcmin$^2$ $YJK_s$ images centred on the new SMC star cluster candidates, where it can be seen the 
environment within which they are inmersed. Note that the 
displayed fields are nearly 7 times larger in radius respect to the clusters' sizes. 
Although the star clusters are centred on the images, the distribution of
field stars along the line of sight as well as the scale of brightness used to produce
the images might lead the reader to confuse them with groups of bright field stars and/or
clouds of gas/dust. We recall that the new star cluster candidates are identified not only as stellar overdenssities
but also from their field star cleaned CMDs.

We estimated reddening values and ages for our 38 new star cluster candidates using the theoretical 
isochrones of \citet{betal12} in the Vegamag system (where, by definition, Vega has a 
magnitude of zero in all filters), corrected by -0.074 mag in $Y$ and -0.003 mag
in $K_s$ to put them on the VMC system \citep{retal15}, to match the cleaned star cluster 
CMDs. We adopted the same distance modulus for all star clusters $(m-M)_o$ = 18.9 $\pm$ 0.1 
mag and $K_s - M_{K_s} = (m-M)_o + 0.372\times E(B-V)$, for $R_V = 3.1$
\citep{cetal89,getal13}, since changes in the distance modulus by an amount equivalent to
the average depth for this SMC region derived 
by \citet[][see their Fig. 8]{retal15}, leads to a smaller age difference than that resulting 
from the isochrones (characterized by the same metallicity) bracketing the observed star cluster 
features in the CMD. We chose isochrones for $Z$ = 0.003 ([Fe/H] = -0.7 dex), which corresponds 
to the mean SMC star cluster metal content for the last $\sim$ 2-3 Gyr \citep{pg13}. Note that 
the $Y-K_s$ colour is not sensitive to metallicity differences smaller than $\Delta$[Fe/H] $\sim$ 0.4 dex, which is adequate given the spread of the stars in the CMDs \citep[see][]{petal15a}. 

We found  in general that isochrones bracketing the derived mean age by $\Delta$log($t$ yr$^{-1}$) = $\pm$0.1 
represent the overall uncertainties associated with the observed dispersion in the cluster
CMDs. Although the dispersion is smaller in some cases, we prefer to retain the former values 
as an upper limit to our error budget.  VMC1, 25 and 37 deserved particular attention.
These object fields have few relatively bright stars with $P \ge$ 50 per cent, so that instead of
deriving
their ages we only could estimate older age limits. As for reddening errors we found a slight correlation 
with the derived $E(B-V)$ values which, in some cases, could be related to the presence of
differential reddening. For star clusters with $E(B-V) \le$ 0.35 mag, the reddening uncertainty 
is estimated as 0.05 mag while for star clusters with larger $E(B-V)$ values, the uncertainty 
reaches 0.10 mag. These reddening uncertainties are thought to mainly represent the overall 
dispersion along the star cluster CMD features, rather than a measure of the total reddening 
spread. Nevertheless, in most of the star clusters the adopted reddening uncertainties fairly
reflect the observed MS broadness, thus implying that only some star clusters show clear effects of
differential reddening. VMC3 presents differential reddening of $\la$ 0.5 mag, and we used the
lower value $E(B-V)$ = 0.7 mag to match the isochrones. Table 1 lists the adopted central  coordinates, radii, the resulting reddening and age values,
and the number of stars with $P \ge$ 75 per cent. The latter is meant to provide
an estimate of the cluster candidate populousness, although some outliers are included. Fig. A2 of the Appendix illustrates the results of 
the isochrone matching for the entire new star cluster candidate sample.

\begin{table}
\caption{Fundamental parameters of new SMC star clusters.}
\label{tab:table1}
\begin{tabular}{@{}lcccccc}\hline
Name & R.A. & Dec. & $r$ &$E(B-V)$ & log($t$) & N$^a$ \\
     & (\degr) & (\degr) & (') & (mag) &    & \\\hline
VMC1 &  11.169& -73.360&    0.15 &    0.90 &   $<$8.4$\pm$0.3& 3\\
VMC2 &  11.468& -73.232&    0.20 &    0.40 &   8.5& 10\\
VMC3 &  11.655& -73.098&    0.40 &    0.70 &   9.0& 55\\
VMC4 &  11.890& -73.158&    0.30 &    0.30 &   8.9& 22\\
VMC5 &  11.895& -73.253&    0.30 &    0.50 &   8.4& 12 \\
VMC6 &  11.982& -73.243&    0.20 &    0.50 &   9.0& 14 \\
VMC7 &  12.045& -73.314&    0.30 &    0.35 &   8.9& 16 \\
VMC8 &  12.090& -73.345&    0.20 &    0.25 &   9.0& 10 \\
VMC9 &  12.106& -73.099&    0.30 &    0.90 &   9.0& 23 \\
VMC10&  12.112& -72.984&    0.40 &    0.25 &   8.1& 25\\
VMC11&  12.133& -73.135&    0.25 &    0.30 &   7.8 & 6\\
VMC12&  12.137& -73.345&    0.25 &    0.40 &   8.7& 13 \\
VMC13&  12.139& -73.392&    0.30 &    0.30 &   8.2& 16 \\
VMC14&  12.146& -73.035&    0.25 &    0.25 &   8.9& 13 \\
VMC15&  12.212& -72.954&    0.30 &    0.25 &   8.9 & 16 \\
VMC16&  12.229& -73.305&    0.25 &    0.20 &   9.0& 11 \\
VMC17&  12.254& -72.913&    0.35 &    0.30 &   8.4& 19 \\
VMC18&  12.267& -73.265&    0.30 &    0.40 &   7.9 & 8\\
VMC19&  12.323& -72.940&    0.40 &    0.35 &   8.5 & 35\\
VMC20&  12.334& -72.995&    0.30 &    0.25 &   8.7& 16 \\
VMC21&  12.338& -72.920&    0.35 &    0.45 &   8.3& 19\\
VMC22&  12.349& -73.165&    0.30 &    0.30 &   8.4& 18 \\
VMC23&  12.427& -73.234&    0.35 &    0.40 &   7.8 & 15\\
VMC24&  12.496& -72.910&    0.25 &    0.20 &   8.7 & 18\\
VMC25&  12.500& -73.408&    0.40 &    0.30 &    $<$7.6$\pm$0.4 & 17\\
VMC26&  12.612& -73.162&    0.40 &    0.20 &   8.1& 18 \\
VMC27&  12.613& -73.311&    0.40 &    0.20 &   8.1& 23 \\
VMC28&  12.650& -73.056&    0.30 &    0.40 &   8.3& 16 \\
VMC29&  12.669& -73.043&    0.40 &    0.30 &   8.2 & 16\\
VMC30&  12.681& -73.261&    0.25 &    0.15 &   8.8& 13 \\
VMC31&  12.681& -73.013&    0.35 &    0.20 &   8.9& 30 \\
VMC32&  12.723& -73.175&    0.30 &    0.15 &   8.1& 12 \\
VMC33&  12.730& -72.915&    0.30 &    0.30 &   8.8& 26\\
VMC34&  12.748& -73.009&    0.40 &    0.25 &   8.1& 27\\
VMC35&  12.770& -73.248&    0.30 &    0.20 &   7.5& 11\\
VMC36&  12.771& -73.085&    0.35 &    0.30 &   8.1& 14 \\
VMC37&  12.793& -73.263&    0.30 &    0.25 &   $<$7.3$\pm$0.5& 12 \\
VMC38&  12.824& -73.180&    0.30 &    0.35 &   8.4& 14 \\
\hline
\end{tabular}

\noindent $^a$ Number of stars within the cluster radius with $P \ge$ 75 per cent.
\end{table}

The Washington cleaned CMDs around the central positions of the new star clusters  
allow us to know their observed features in the optical wavelength regime and 
consequently, to compare them to those in the near-infrared. By doing this exercise,
we found that most of the star clusters are detectable from a sound Washington
CMD analysis, once the star field signature is properly filtered. We used the derived
reddenings and ages of Table 1, as well as a distance modulus  $(m-M)_o$ = 18.9 and
a metallicity of $Z$ = 0.003, to match theoretical isochrones \citep{betal12} to the
cleaned $T_1$ versus $C-T_1$ CMDs. The theoretical isochrones were properly shifted
by the corresponding $E(C - T_1) = 1.97E(B - V)$ colour excesses and by the SMC 
distance modulus $T_1 = M_{T_1} + (m-M)_o + 2.62E(B-V)$ \citep{gs99}. We confirmed our 
estimated values of $E(B-V)$ and log($t$ yr$^{-1}$), as illustrated in Fig. 3. As can 
be seen, the Washington photometry is slightly deeper than the VMC one and possibly, 
due to a relatively low reddening along the line-of-sight, the generally blue 
integrated light nature of young/moderate age star clusters makes their MS appear 
reasonably well-populated. At the top of Fig. 3 we show the deepest $C$, $T_1$ and 
$K_s$ images (from left ot right) of the star cluster field. Note that the star cluster 
clearly stands out at the near-infrared image, while it is hardly possible to 
recognize it from the optical one without the help of the corresponding CMD analysis.

\begin{figure*}
\fbox{\includegraphics[width=\textwidth]{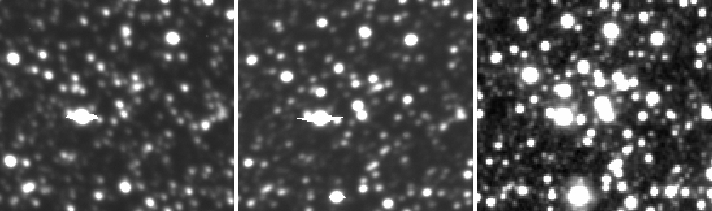}} \\
\includegraphics[width=\textwidth]{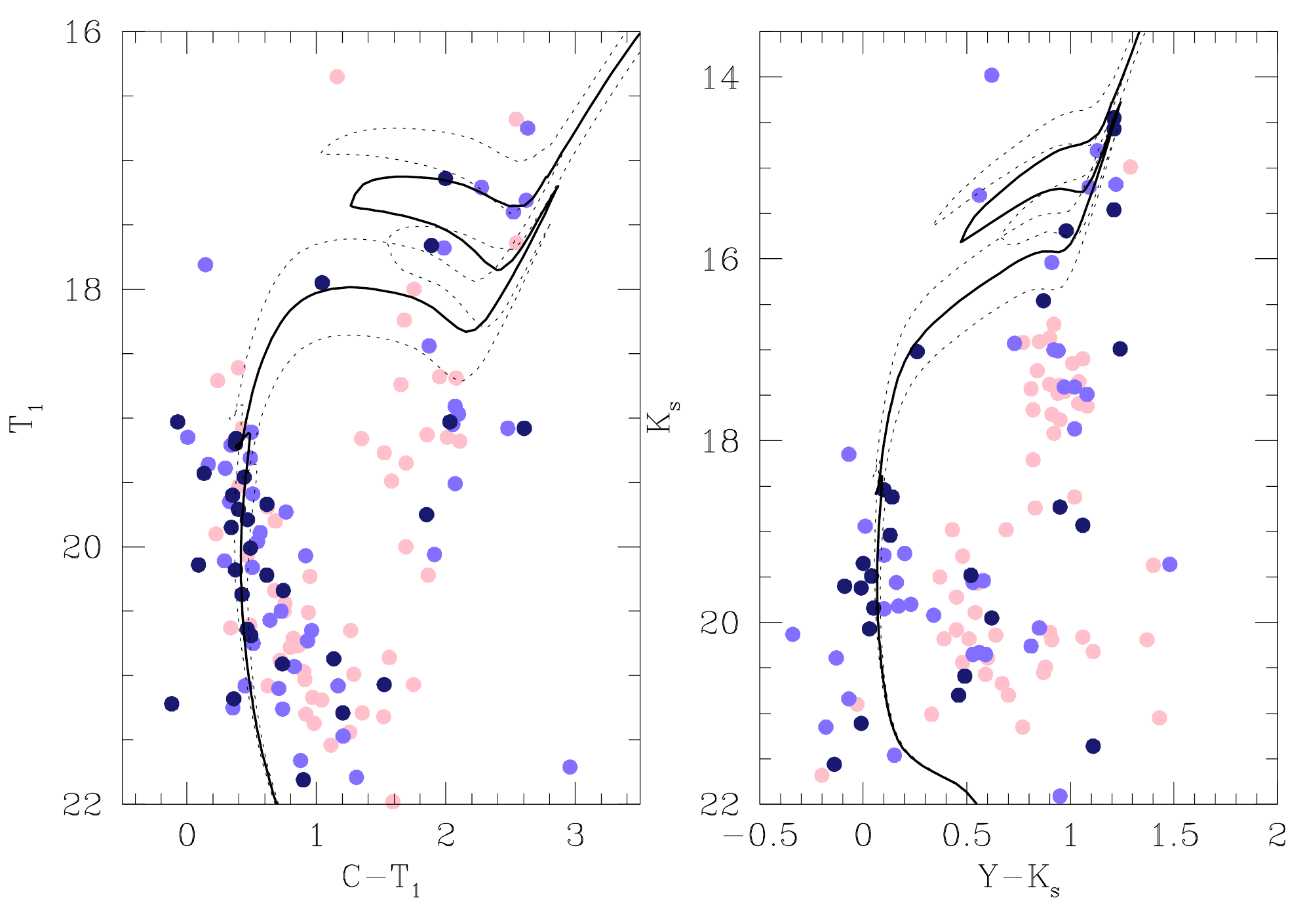}
\caption{{\it Top:} $C$, $T_1$ and $K_s$ images (from left to right) centred on the new 
star cluster candidate VMC17. The images' sides are equal to the star cluster diameter (see
Table 1). North is up and East to the left. 
{\it Bottom:} Washington and VMC CMDs illustrating the performance of 
the cleaning procedure, where we plotted all the measured stars located within
the cluster radius. The pink, light and dark blue filled circles in the bottom 
panels represent stars with cluster membership probabilities $P \le$ 25 per cent, 
$P =$ 50 per cent, and $P \ge$ 75 per cent, respectively.
Three isochrones from \citet{betal12} for log($t$ yr$^{-1}$)
  = 8.3, 8.4, and 8.5 and $Z$ = 0.003 are also superimposed.}
\end{figure*}

Three cluster candidates (VMC\,1, 3 and 9), on the other hand, could not be identified neither by a
visual inspection of their $C,T_1$ images nor from the respective cleaned CMDs. 
We found that all of them are affected by reddening higher than $E(B-V)$ $\sim$ 0.6 mag,
so that we conclude that this could be the reason for making them undetectable from
optical photometry. Fig. 4 illustrates this situation, in which the 1 Gyr old
star cluster  candidate VMC9 appears hidden at optical wavelengths. This
example, in turn, brings support to our original motivation for this search, in the
sense that the VMC survey should have imaged those star clusters that are either behind or embedded within clouds in the Magellanic system. According to \citet{retal15}, who 
produced an $A_V$ extinction map for the SMC main body, our
pilot field is the most reddened in the galaxy, with best-fitting $E(B-V)$ values 
larger than 0.20 mag. Their $A_V$ extinction map was built  by running the StarFISH
SFH-recovery software over a wide-enough grid of distance
modulus and extinction values until the code
returns not only the best-fitting coefficients for the SFH, but
also the best-fitting ($(m-M )_o, A_V$) pair (see their Fig. 11).

\begin{figure*}
\fbox{\includegraphics[width=\textwidth]{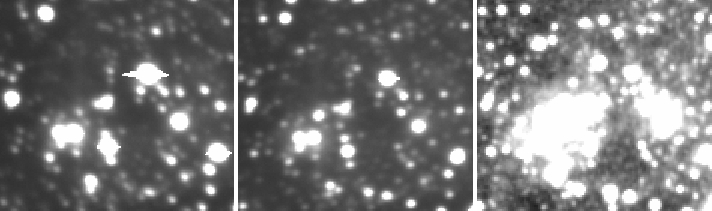}} \\
\includegraphics[width=\textwidth]{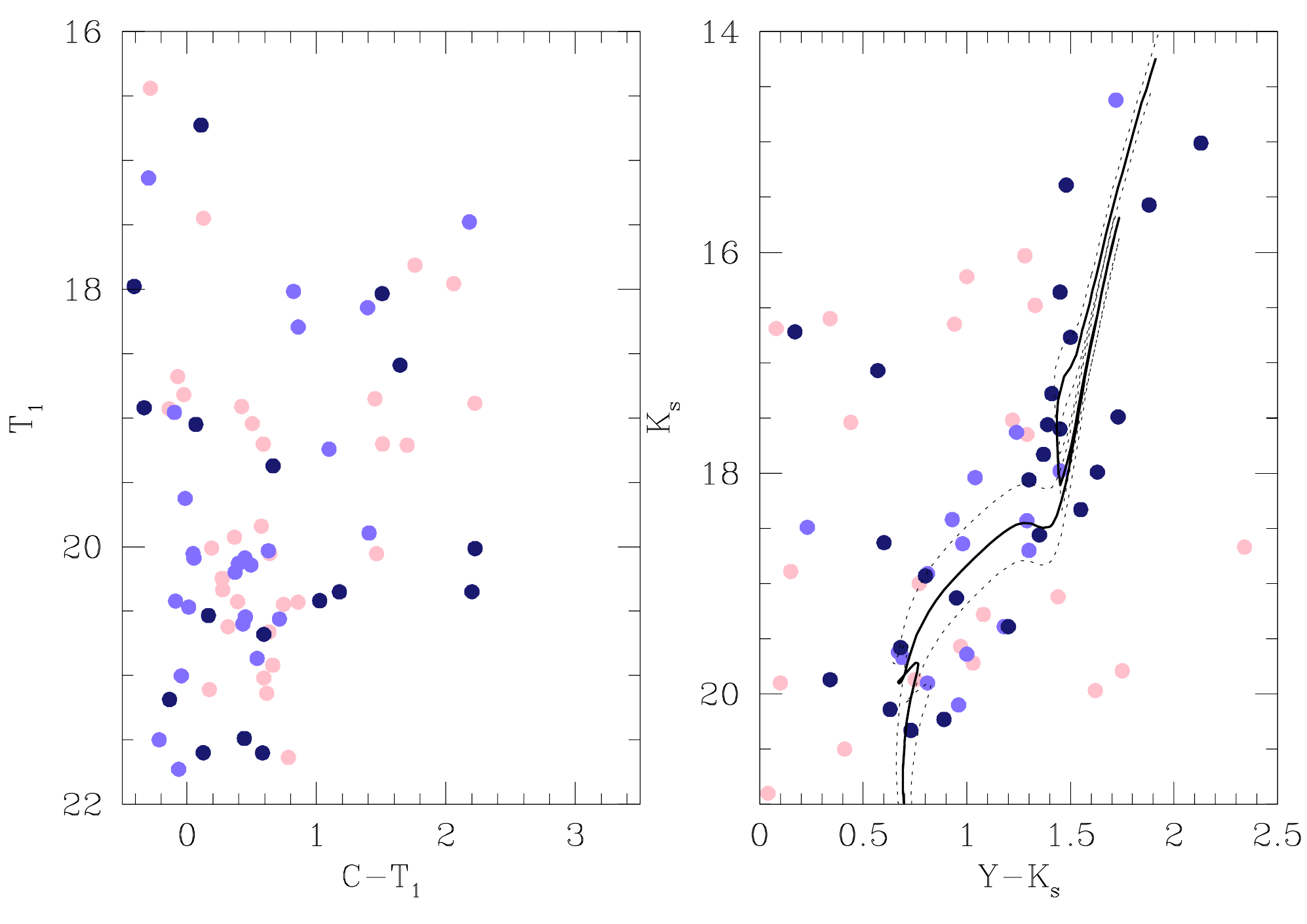}
\caption{Same as Fig. 3 for the new star cluster candidate VMC9. Three isochrones from 
\citet{betal12} for log($t$ yr$^{-1}$)  = 8.9, 9.0, and 9.1 and $Z$ = 0.003 are 
also superimposed.}
\end{figure*}

\section{conclusions}

By using the near-infrared VMC data set we performed a search for new star 
clusters in the South-West side of the SMC bar, where the star field is the
densest and hightest reddened region in the galaxy. The search was motivated
by the fact that star clusters not seen in the visible could be dwelling such
regions, as judged by the comparison of optical and VMC images.

The devised procedure for the star cluster search relies on astrophysical information taken 
from the VMC PSF photometric catalogues for the known star clusters located 
in a pilot field of $\sim$ 0.4 deg$^2$. From the learned distribution of the dimensions
and the observed stellar densities of known star clusters, we designed a strategy for 
finding new ones that consisted in using {\it Gaussian} and {\it tophat} KDEs for three 
different bandwidths. After perfoming six different runs over an amount of 
358578 stars with measurements in the three $YJK_s$ filters, we detected 143 new star cluster 
candidates, within a similar range of radius and stellar density to the previously 
catalogued star cluster sample.

We applied a subtraction procedure developed by \citet{pb12} to statistically clean the 
star cluster CMDs from field star contamination in order to disentangle star cluster features from 
those belonging to their surrounding fields. The employed technique makes use of variable 
cells in order to reproduce the field CMD as closely as possible. As a result 38 objects
of relatively small size - on average $\sim$ 0.3 arcmin in radius - resulted to have 
near-infrared CMD features which resemble those of star clusters of young to moderate 
intermediate age (log($t$ yr$^{-1}$) $\sim$ 7.5-9.0). The new star cluster candidates represent 
$\approx$ a 55$\%$ increase on the known 
star cluster population, which is particularly significant in the light of the current
debates about the star cluster formation rate, the effectiveness of star cluster dissolution 
processes, etc.

From matching theoretical isochrones computed for the VISTA system to 
the cleaned star cluster CMDs we estimated reddenings and ages. When adjusting a subset of 
isochrones we took into account the SMC distance modulus and the mean SMC star cluster 
metal content for the last $\sim$ 2-3 Gyr. The derived mean $E(B-V)$ colour excesses are
in between 0.15 mag and 0.90 mag, while their ages are in the range  7.3 $\le$ 
log($t$ yr$^{-1}$) $\le$ 9.0. This new star cluster candidate sample will be part of the cluster
database that the VMC survey will produce in order to homogeneously study the overall star
cluster formation history throughout the Magellanic system.

When comparing the cleaned star cluster CMDs obtainted from $CT_1$ Washington photometry,
by employing the same star field decontamination method mentioned above, to those from the 
VMC survey, we found that most of the new star clusters are detectable, and confirm our 
estimates of $E(B-V)$ and log($t$ yr$^{-1}$). Likewise, it is worth mentioning that the
star clusters are clearly visible in the deepest $K_s$ images, whereas it is hardly possible to 
recongnize them from the optical ones without the help of the corresponding CMD analysis.
Whenever the star clusters are affected by reddening higher than $E(B-V)$ $\ge$ 0.6 mag, they
could not be recognized from the analysis of Washington photometry, thus supporting
the crucial complementary role of near-infrared bands surveys.

\section*{Acknowledgements}
We thank the Cambridge Astronomy Survey Unit (CASU) and the Wide-Field
Astronomy Unit (WFAU) in Edinburgh for providing calibrated data
products under the support of the Science and Technology Facilities
Council (STFC) in the UK. This research has made use of the SIMBAD
data base, operated at CDS, Strasbourg, France.  We also thank
Gabriel Perren, Omar Silvestro and Roberto Cattaneo for providing 
computer-programing support during the developement
of this work. MRC acknowledges support from the UK's STFC [grant number ST/M001008/1] 
and from the German Academic Exchange Service (DAAD).
 We thank the anonymous referee whose thorough comments and suggestions
allowed us to improve the manuscript.




\bibliographystyle{mnras}

\input{paper.bbl}




\appendix
\section{New SMC star clusters}

\begin{figure*}
\fbox{\includegraphics[width=4.0cm]{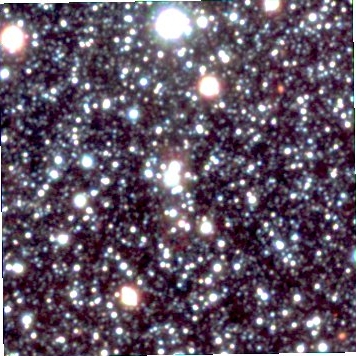}}
\fbox{\includegraphics[width=4.0cm]{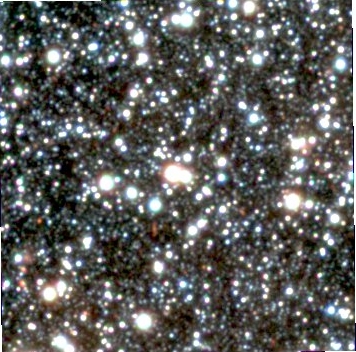}}
\fbox{\includegraphics[width=4.0cm]{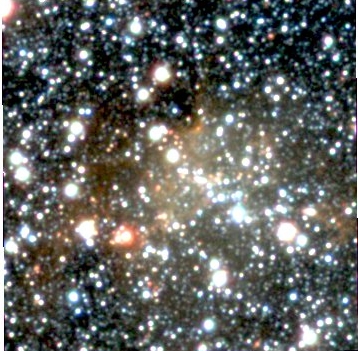}}
\fbox{\includegraphics[width=4.0cm]{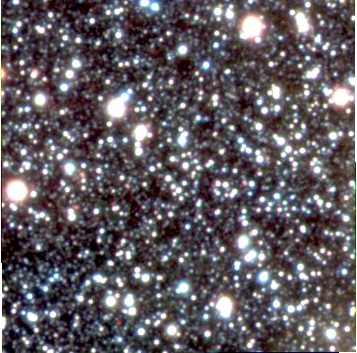}} \\
\framebox[4.25cm]{VMC1}
\framebox[4.25cm]{VMC2}
\framebox[4.25cm]{VMC3}
\framebox[4.25cm]{VMC4}\\
\fbox{\includegraphics[width=4.0cm]{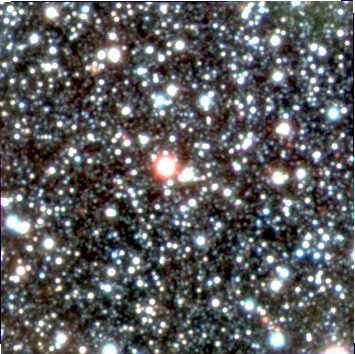}}
\fbox{\includegraphics[width=4.0cm]{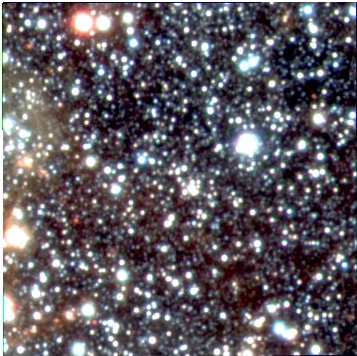}}
\fbox{\includegraphics[width=4.0cm]{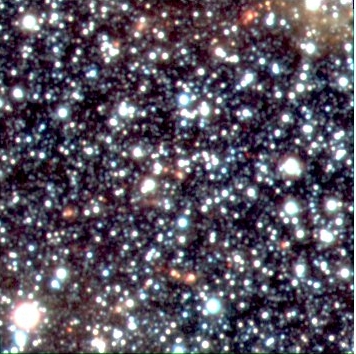}}
\fbox{\includegraphics[width=4.0cm]{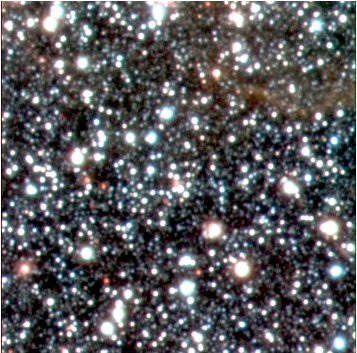}} \\
\framebox[4.25cm]{VMC5}
\framebox[4.25cm]{VMC6}
\framebox[4.25cm]{VMC7}
\framebox[4.25cm]{VMC8}\\
\fbox{\includegraphics[width=4.0cm]{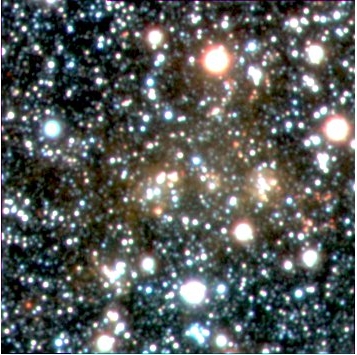}}
\fbox{\includegraphics[width=4.0cm]{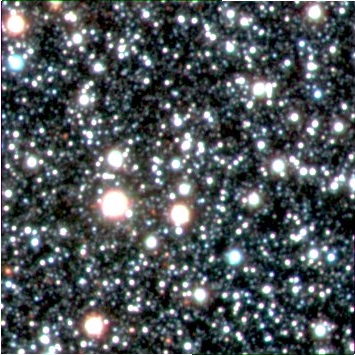}}
\fbox{\includegraphics[width=4.0cm]{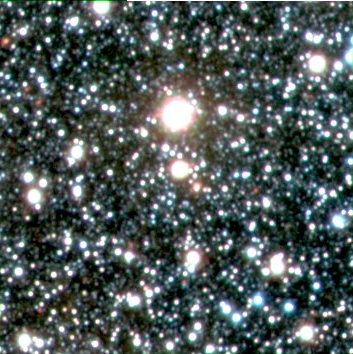}}
\fbox{\includegraphics[width=4.0cm]{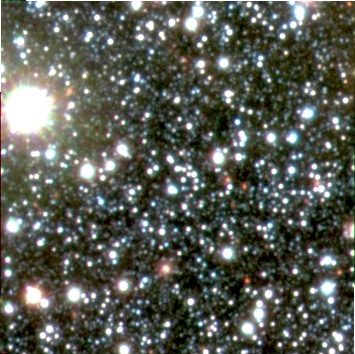}} \\
\framebox[4.25cm]{VMC9}
\framebox[4.25cm]{VMC10}
\framebox[4.25cm]{VMC11}
\framebox[4.25cm]{VMC12}\\
\fbox{\includegraphics[width=4.0cm]{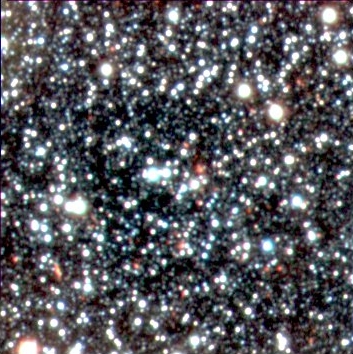}}
\fbox{\includegraphics[width=4.0cm]{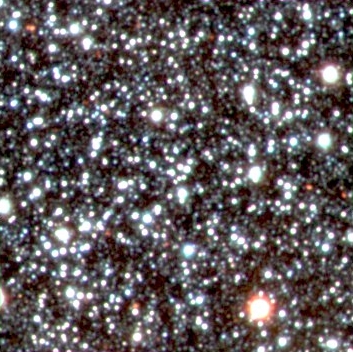}}
\fbox{\includegraphics[width=4.0cm]{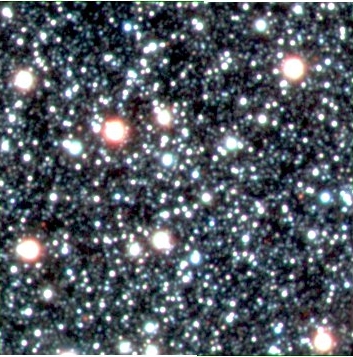}}
\fbox{\includegraphics[width=4.0cm]{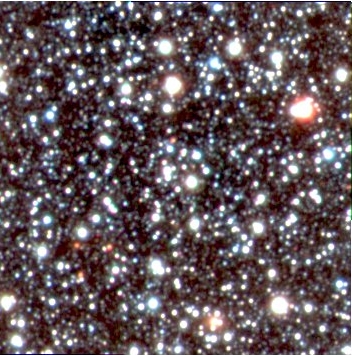}} \\
\framebox[4.25cm]{VMC13}
\framebox[4.25cm]{VMC14}
\framebox[4.25cm]{VMC15}
\framebox[4.25cm]{VMC16}\\
\caption{2$\times$2 arcmin$^2$ $YJK_s$ images centred on the new
SMC star clusters.  North is up and East to the left. Note that the star clusters' radii are between 0.15 and 0.40 arcmin.}
\end{figure*}

\setcounter{figure}{0}
\begin{figure*}
\fbox{\includegraphics[width=4.0cm]{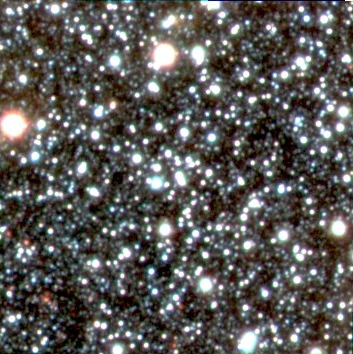}}
\fbox{\includegraphics[width=4.0cm]{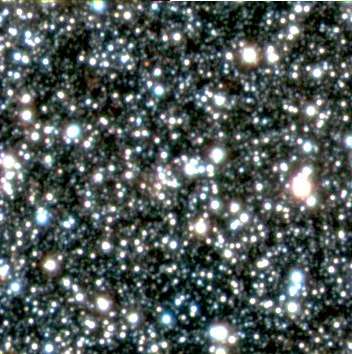}}
\fbox{\includegraphics[width=4.0cm]{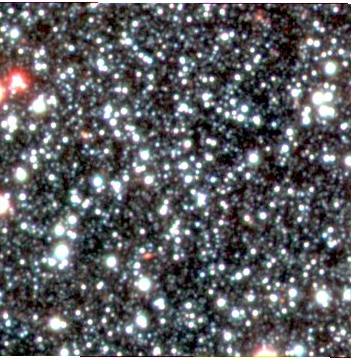}}
\fbox{\includegraphics[width=4.0cm]{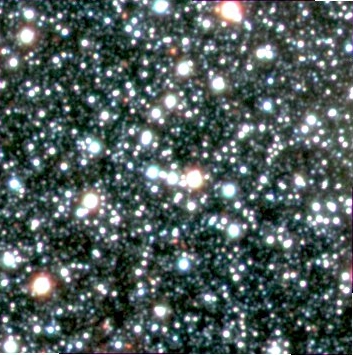}} \\
\framebox[4.25cm]{VMC17}
\framebox[4.25cm]{VMC18}
\framebox[4.25cm]{VMC19}
\framebox[4.25cm]{VMC20}\\
\fbox{\includegraphics[width=4.0cm]{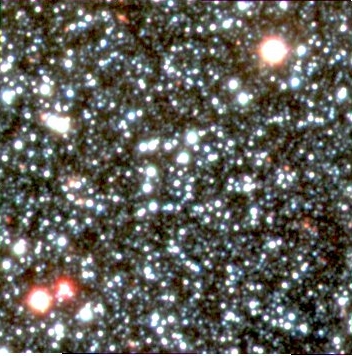}}
\fbox{\includegraphics[width=4.0cm]{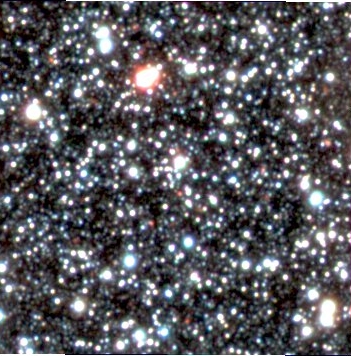}}
\fbox{\includegraphics[width=4.0cm]{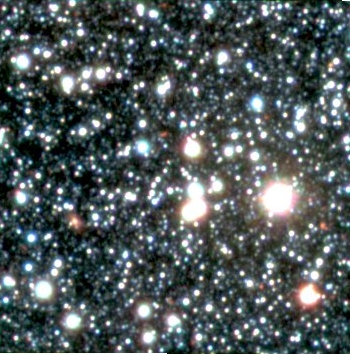}}
\fbox{\includegraphics[width=4.0cm]{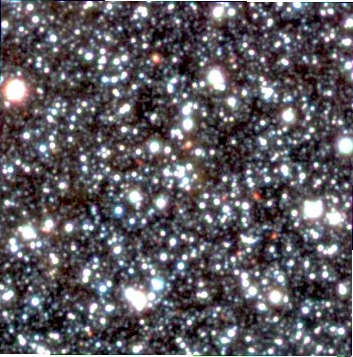}} \\
\framebox[4.25cm]{VMC21}
\framebox[4.25cm]{VMC22}
\framebox[4.25cm]{VMC23}
\framebox[4.25cm]{VMC24}\\
\fbox{\includegraphics[width=4.0cm]{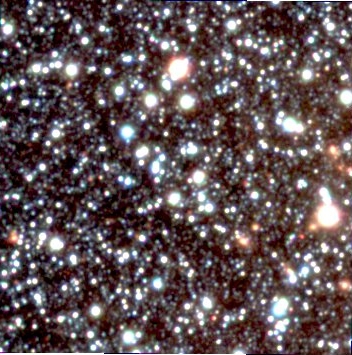}}
\fbox{\includegraphics[width=4.0cm]{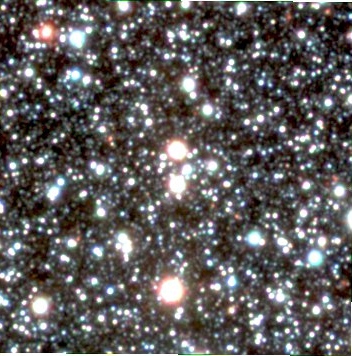}}
\fbox{\includegraphics[width=4.0cm]{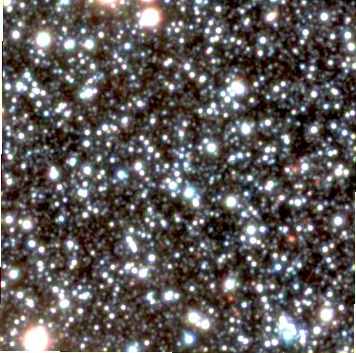}}
\fbox{\includegraphics[width=4.0cm]{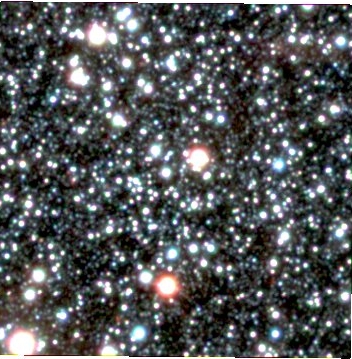}} \\
\framebox[4.25cm]{VMC25}
\framebox[4.25cm]{VMC26}
\framebox[4.25cm]{VMC27}
\framebox[4.25cm]{VMC28}\\
\fbox{\includegraphics[width=4.0cm]{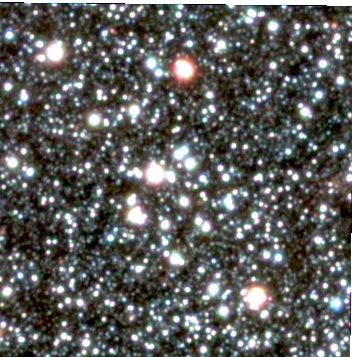}}
\fbox{\includegraphics[width=4.0cm]{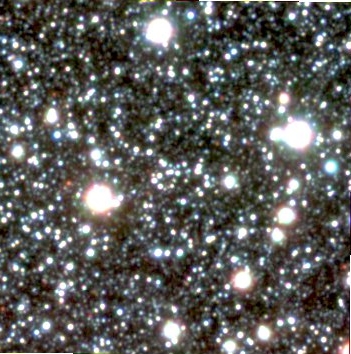}}
\fbox{\includegraphics[width=4.0cm]{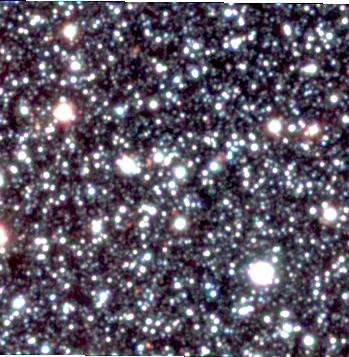}}
\fbox{\includegraphics[width=4.0cm]{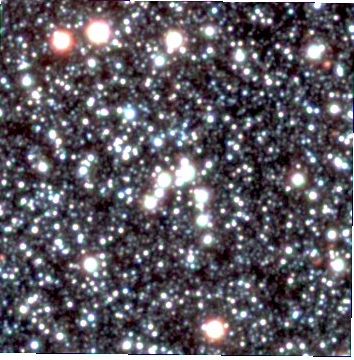}} \\
\framebox[4.25cm]{VMC29}
\framebox[4.25cm]{VMC30}
\framebox[4.25cm]{VMC31}
\framebox[4.25cm]{VMC32}\\
\caption{continued.}
\end{figure*}

\setcounter{figure}{0}
\begin{figure*}
\fbox{\includegraphics[width=4.0cm]{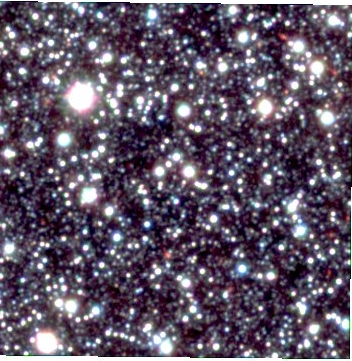}}
\fbox{\includegraphics[width=4.0cm]{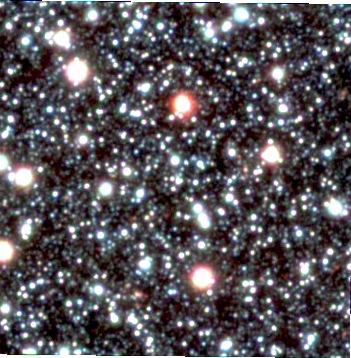}}
\fbox{\includegraphics[width=4.0cm]{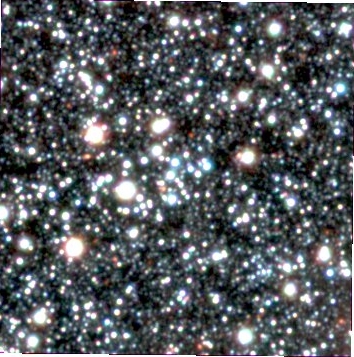}}
\fbox{\includegraphics[width=4.0cm]{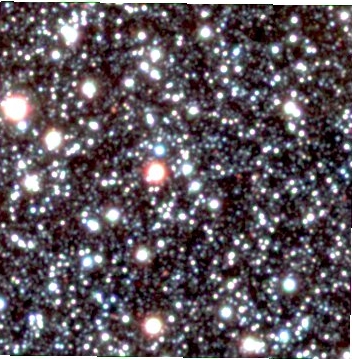}} \\
\framebox[4.25cm]{VMC33}
\framebox[4.25cm]{VMC34}
\framebox[4.25cm]{VMC35}
\framebox[4.25cm]{VMC36}\\
\fbox{\includegraphics[width=4.0cm]{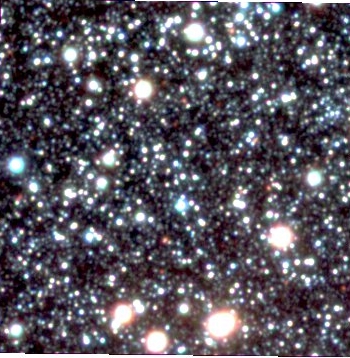}}
\fbox{\includegraphics[width=4.0cm]{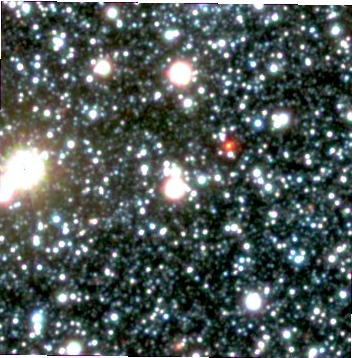}}\\
\framebox[4.25cm]{VMC37}
\framebox[4.25cm]{VMC38}\\
\caption{continued.}
\end{figure*}

\begin{figure*}
	\includegraphics[width=\textwidth]{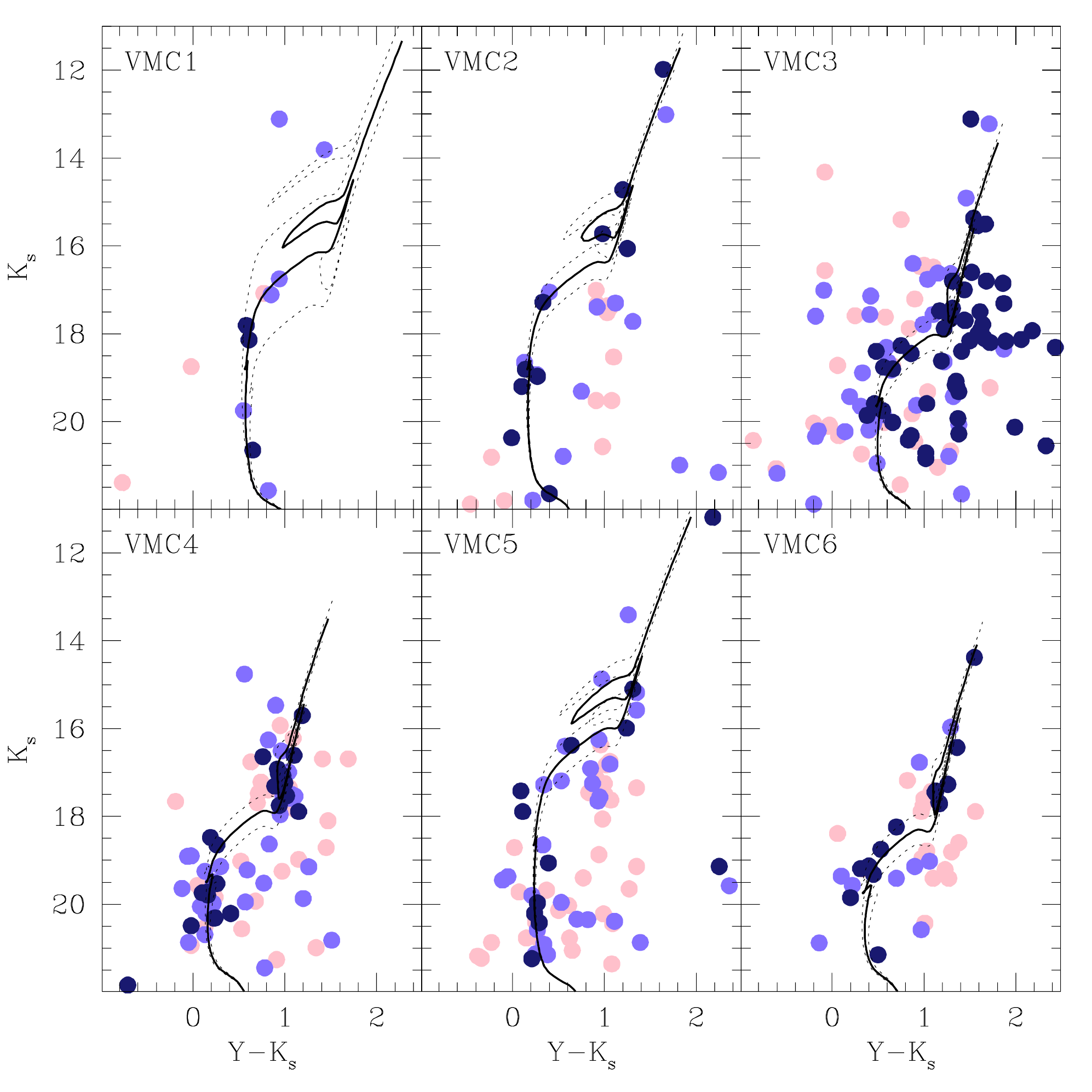}
    \caption{$K_s$ versus $Y-K_s$ CMD of stars within the cluster radius. 
Colour-coded symbols
  represent stars that statistically belong to the field ($P \le$
  25\%, pink), stars that might belong to either the field or the
  cluster ($P =$ 50\%, light blue), and stars that predominantly
  populate the cluster region ($P \ge$ 75\%, dark blue). Three
  isochrones from \citet{betal12} for log($t$ yr$^{-1}$) and log($t$
  yr$^{-1}$) $\pm$ 0.1 are also superimposed.  }
    \label{fig:fig5a}
\end{figure*}

\setcounter{figure}{1}
\begin{figure*}
	\includegraphics[width=\textwidth]{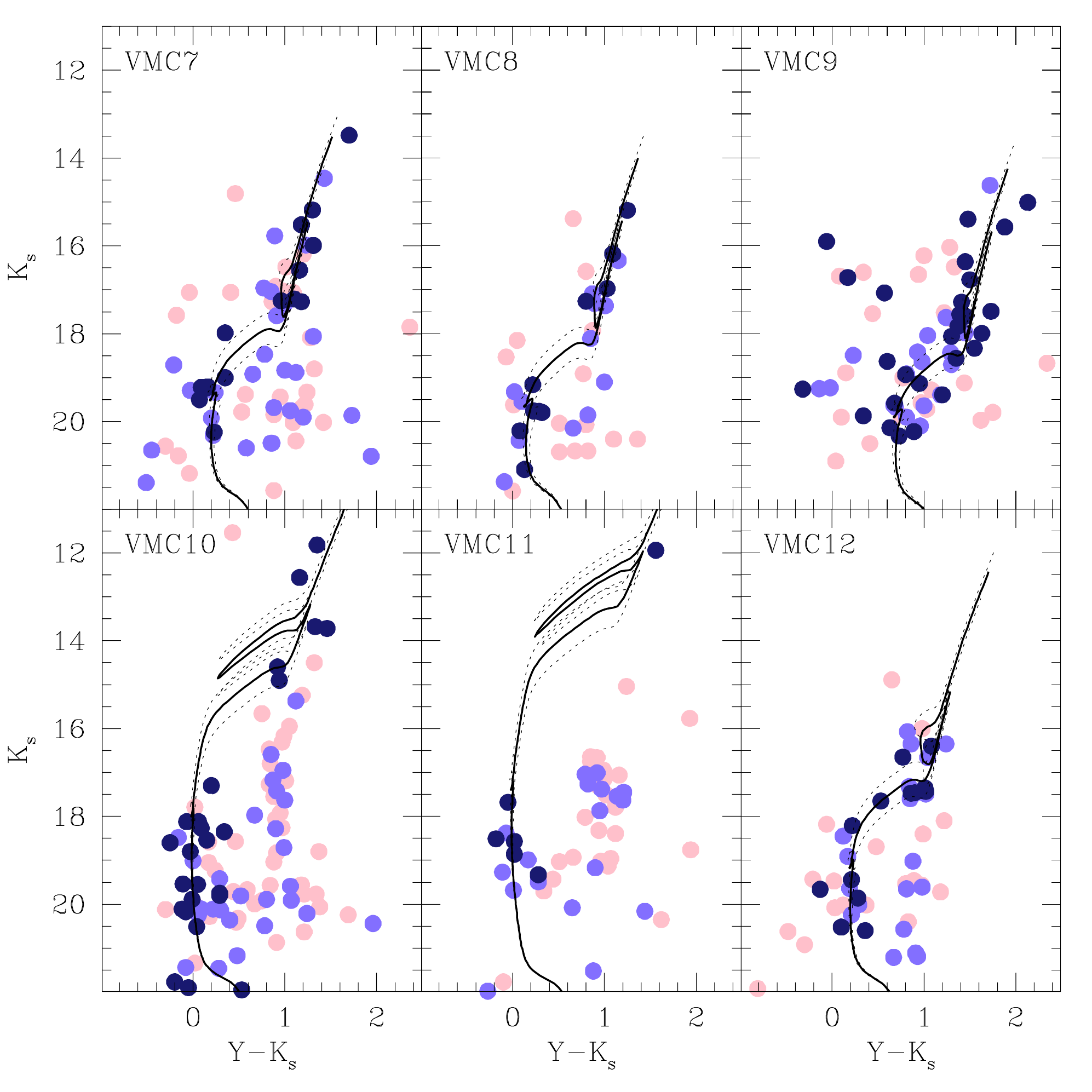}
    \caption{continued.}
    \label{fig:fig5b}
\end{figure*}

\setcounter{figure}{1}
\begin{figure*}
	\includegraphics[width=\textwidth]{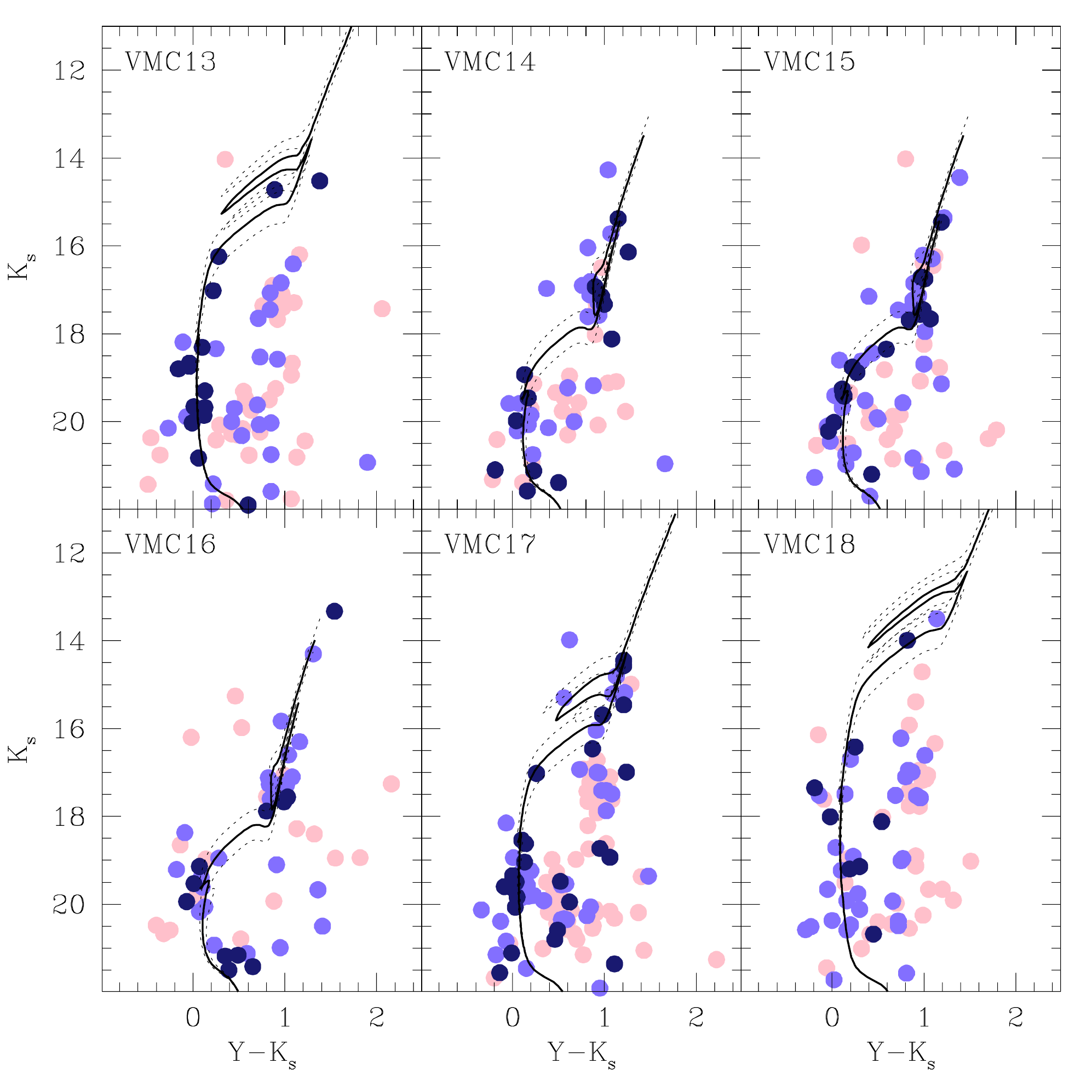}
    \caption{continued.}
    \label{fig:fig5c}
\end{figure*}

\setcounter{figure}{1}
\begin{figure*}
	\includegraphics[width=\textwidth]{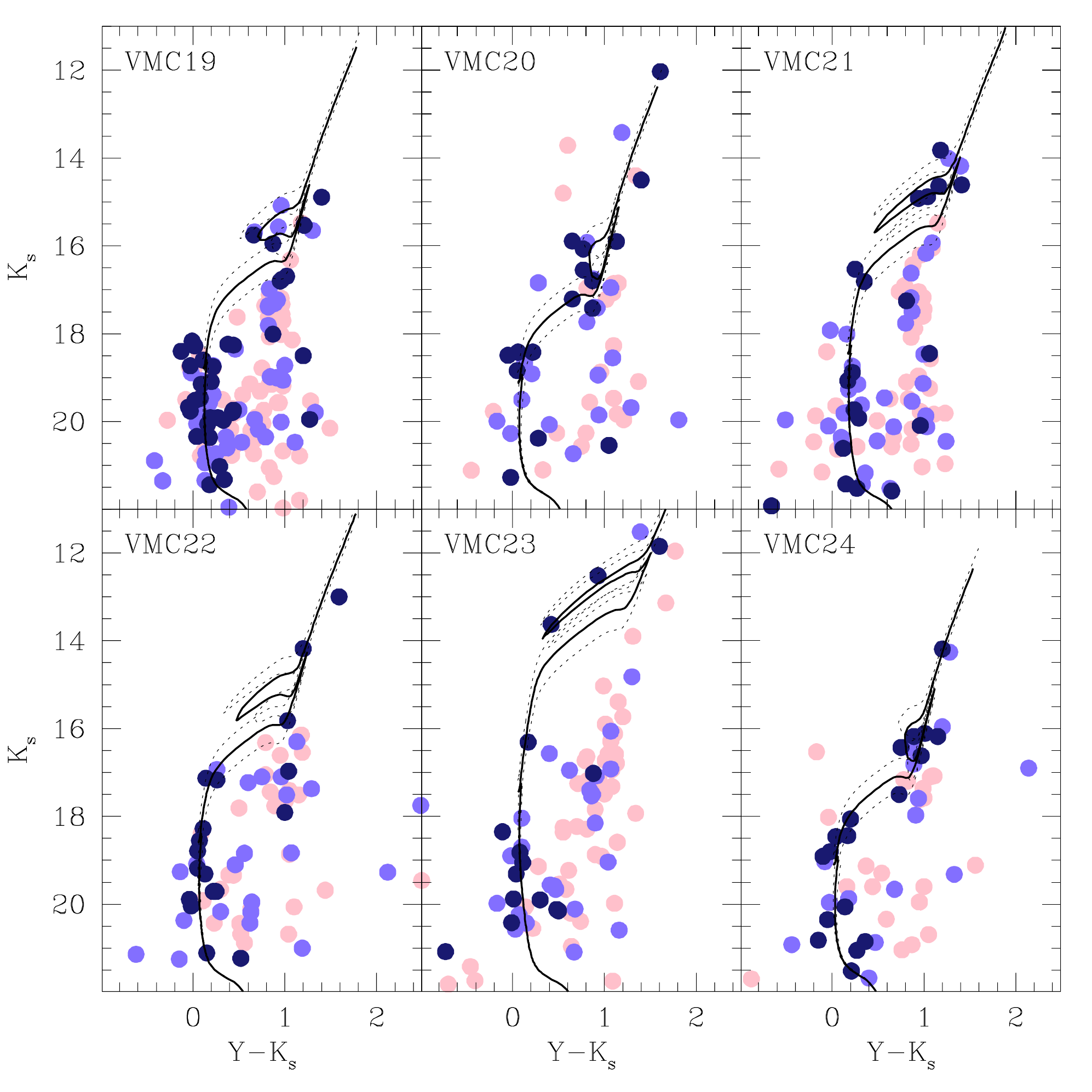}
    \caption{continued.}
    \label{fig:fig5d}
\end{figure*}

\setcounter{figure}{1}
\begin{figure*}
	\includegraphics[width=\textwidth]{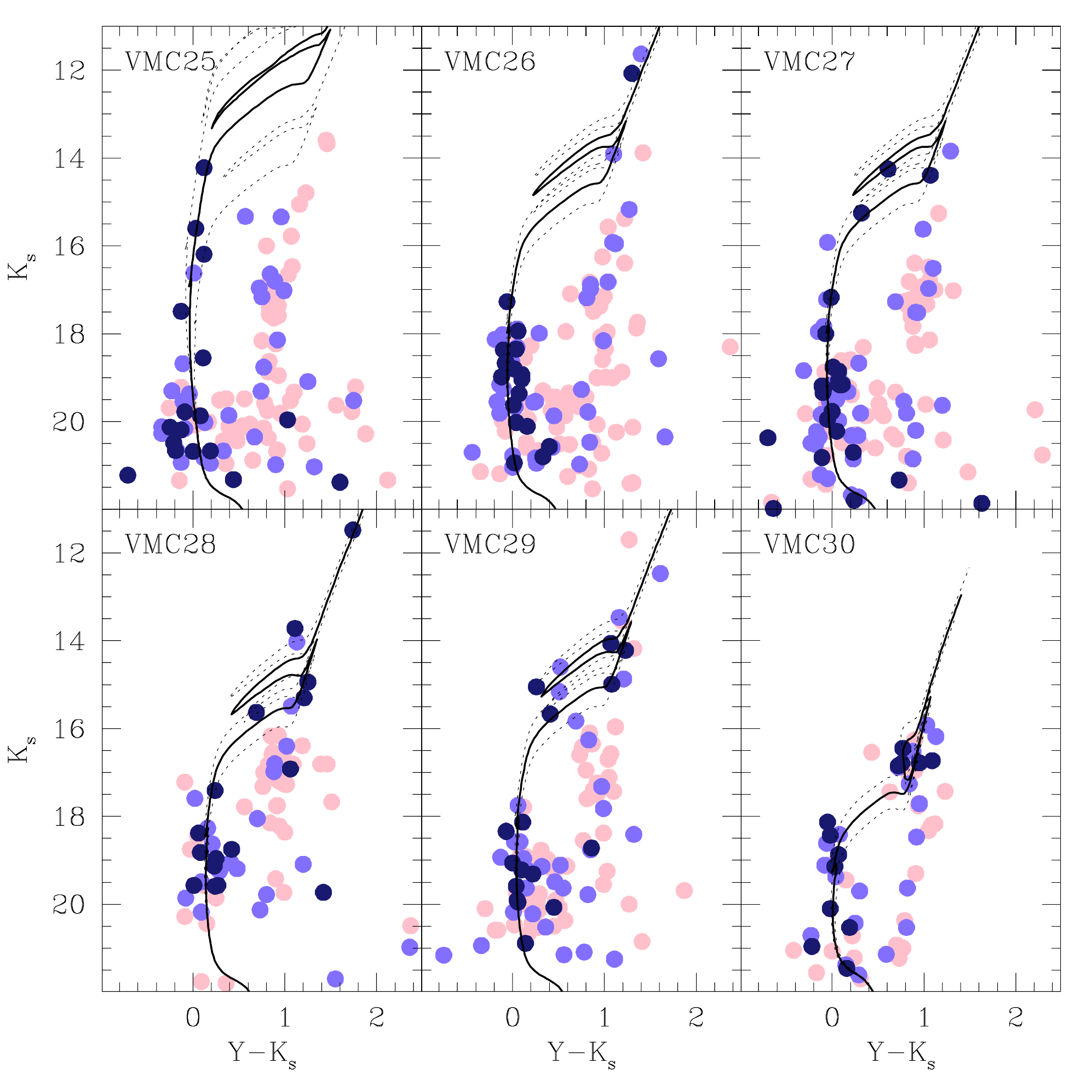}
    \caption{continued.}
    \label{fig:fig5e}
\end{figure*}

\setcounter{figure}{1}
\begin{figure*}
	\includegraphics[width=\textwidth]{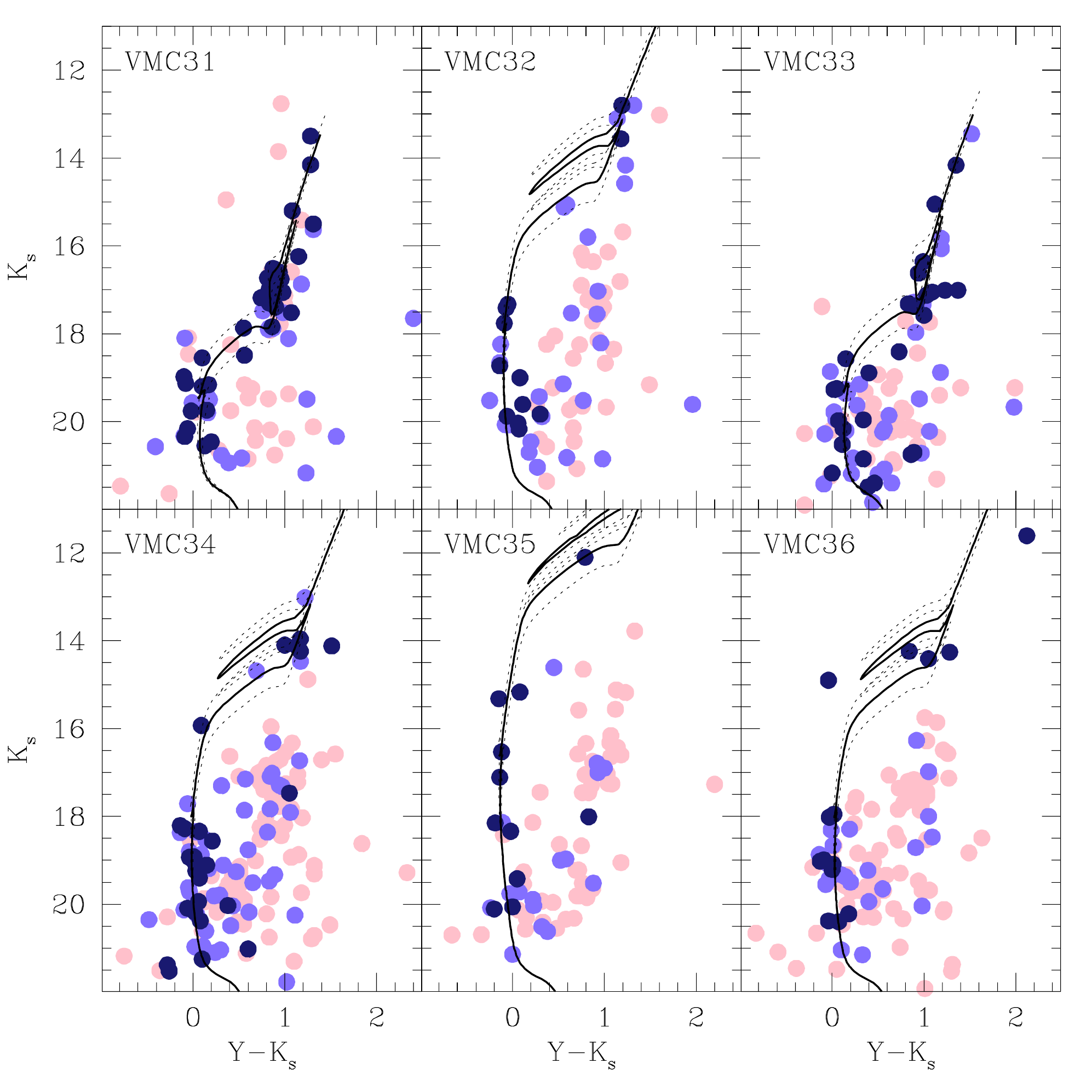}
    \caption{continued.}
    \label{fig:fig5f}
\end{figure*}

\setcounter{figure}{1}
\begin{figure*}
	\includegraphics[width=\textwidth]{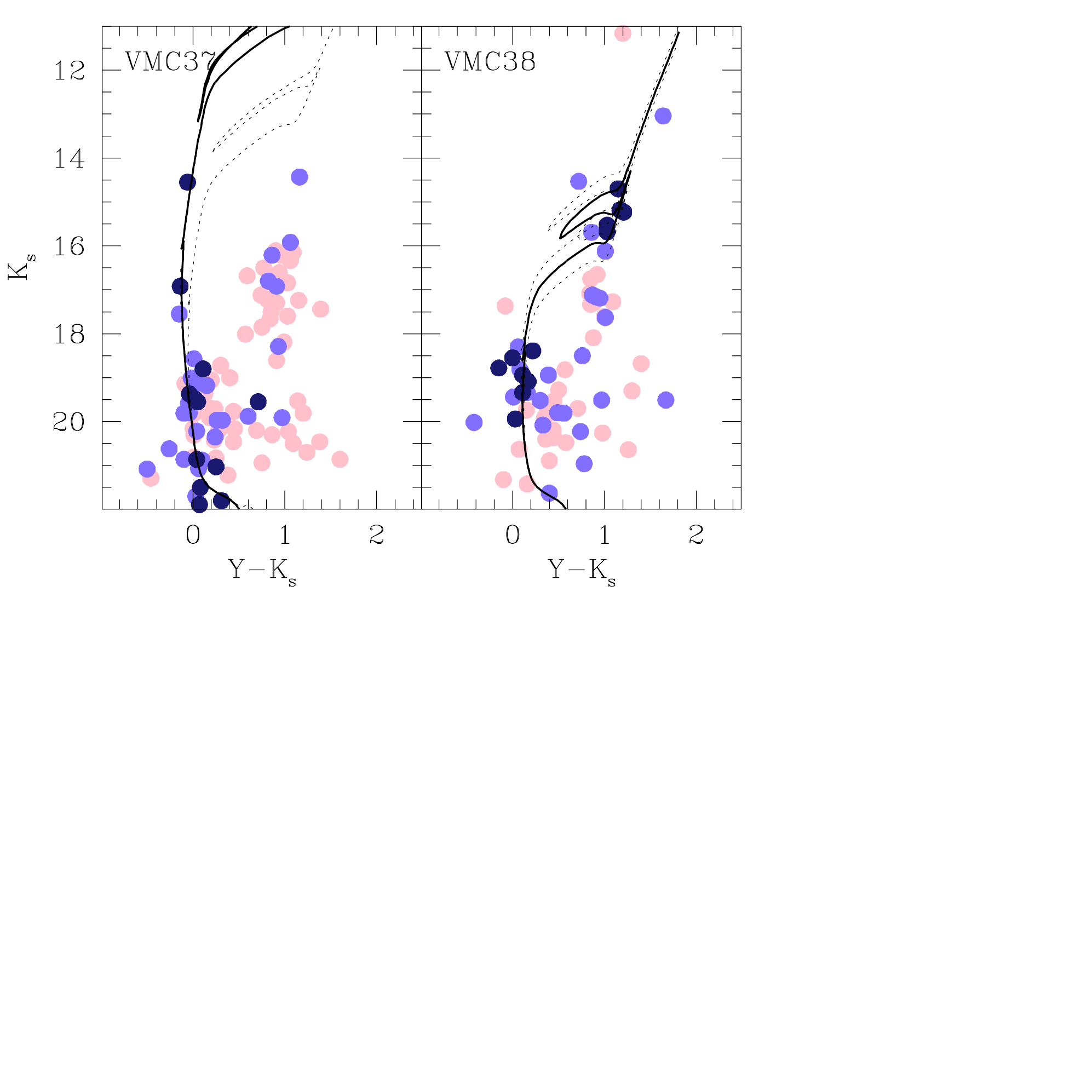}
    \caption{continued.}
    \label{fig:fig5g}
\end{figure*}



\bsp	
\label{lastpage}


\end{document}